\documentclass{aa}
\usepackage{graphicx}
\usepackage{txfonts}
\usepackage{natbib}
\usepackage{hyperref} 
\usepackage{dsfont}
\hypersetup{colorlinks=true,linkcolor=blue,citecolor=blue,urlcolor=blue}
\usepackage{orcidlink}
\usepackage{lscape}

\begin{document}






\title{A Census of Sun's Ancestors and their Contributions to the Solar System Chemical Composition}

\author { F. Fiore  \thanks {email to: FRANCESCA.FIORE2@studenti.units.it} \inst{1}, F. Matteucci \orcidlink{0000-0001-7067-2302} \inst{1,2,3} ,  E. Spitoni \orcidlink{0000-0001-9715-5727} \inst{2}, M. Molero \orcidlink{0000-0002-8854-6547}\inst{4,2}, P. Salucci \orcidlink{0000-0002-5476-2954} \inst{5,3}, D. Romano   \orcidlink{0000-0002-0845-6171} \inst{6} \and A. Vasini \orcidlink{0009-0007-0961-0429} \inst{1,2}
 }
\institute{ Dipartimento di Fisica, Sezione di Astronomia,
  Universit\`a di Trieste, Via G.~B. Tiepolo 11, I-34143 Trieste,
  Italy \and I.N.A.F. Osservatorio Astronomico di Trieste, via G.B. Tiepolo
 11, 34131, Trieste, Italy  
   \and  I.N.F.N. Sezione di Trieste, via Valerio 2, 34134 Trieste, Italy
\and
         Institut f\"ur Kernphysik, Technische Universit\"at Darmstadt, Schlossgartenstr. 2, Darmstadt 64289, Germany
         \and SISSA-International School for Advanced Studies, Via Bonomea 265, 34136 Trieste, Italy
 \and
          INAF, Osservatorio di Astrofisica e Scienza dello Spazio, Via Gobetti 93/3, I-40129 Bologna, Italy
         }

 \date{Received xxxx / Accepted xxxx}

\abstract {
In this work we compute the rates and numbers of different types of stars  and phenomena (supernovae, novae, white dwarfs, merging neutron stars, black holes) that contributed to the chemical composition of the Solar System.  During the Big Bang only light elements formed, while all the heavy ones, from carbon to uranium and beyond, were created inside stars. Stars die and restore the newly formed elements into the interstellar gas. This process is called "chemical evolution".
In particular, we analyse the death rates of stars of all masses, dying either quiescently or explosively. These rates and total star numbers are computed in the context of a revised version of the two-infall model for the chemical evolution of the Milky Way, which reproduces fairly well  the observed abundance patterns of several chemical species, the global solar metallicity, and the current gas, stellar, and total surface mass densities.  We compute also the total number of stars ever born and still alive as well as the number of stars born up to the formation of the Solar System with a mass and metallicity like the Sun. This latter number will account for all the possible existing Solar Systems which can host life in the solar vicinity.
We conclude that, among all the stars (from 0.8 to 100 M$_{\odot}$) born and died from the beginning up to the Solar System formation epoch, which contributed to its chemical composition, 93.00\% are represented by  stars dying as single white dwarfs (without interacting significantly with a companion star) and originating in the mass range 0.8-8 M$_{\odot}$, while 5.24\% are neutron stars and 0.73\% are black holes, both originating from supernovae core-collapse (M > 8 M$_{\odot}$); 0.64\%  are Type Ia supernovae and 0.40\% are nova systems, both originating from the same mass range as the white dwarfs. The number of stars similar to the Sun  born from the beginning up to the Solar System formation, with metallicity  in the range 12+log(Fe/H)= 7.50 $\pm$ 0.04 dex is  $ \sim 31 \cdot$ 10$^{7}$, and in particular our Sun is the $\sim 2.61 \cdot$  10$^7$-th star of this kind.}


\titlerunning{A Census of Sun’s Ancestors}

\authorrunning{Fiore et al.}

\maketitle

\section{Introduction}
\label{sec: introduction}

Stars are born, live and die. During their lives they produce new chemical elements starting from  H and He, in particular they form all the elements from ${}^{12}$C to Uranium and beyond.
They eject newly formed elements, both by stellar winds and through supernovae (SNe) explosions, thus increasing their abundance in the interstellar medium (ISM). This process is known as galactic chemical evolution and it is responsible for the chemical composition of the Solar System, that was born 4.6 Gyr ago (e.g., \citealp{Bouvier2010}).
In order to study chemical evolution we need to build detailed models including several physical ingredients, such as star formation rate, initial mass function, stellar nucleosynthesis and gas flows.
In the literature there are many of such models but very few are detailed enough to follow the evolution of many chemical species  and to take into account all the necessary stellar sources. In particular, massive stars ($M>8M_{\odot}$) ending their lives as core-collapse supernovae (CC-SNe) are responsible for the production of $\alpha$-elements (e.g. O, Ne, Mg, Ca, Si, Ti) as well as r-process elements (also by means of merging neutron stars), while low- and intermediate-mass stars ($0.8\le M/M_{\odot} \le8$) produce C, N and heavy s-process elements (e.g. Ba, Y, La). Supernovae Type Ia (exploding white dwarfs in binary systems) are responsible for the production of most of Fe and novae, also originating from white dwarfs in binary systems, are not negligible producers of CNO isotopes as well as $^{7}$Li. These models relax the instantaneous recycling approximation and compute in detail the rates of SNe of all types, novae, and merging neutron stars. Taking into account the stellar lifetimes is fundamental in order to correctly predict the abundances of the elements and their ratios. Clearly, the stellar yields as functions of the stellar mass and  metallicity represent one of the most important ingredients of such models, together with the stellar birthrate function (star formation rate and initial mass function) and possible gas flows (see \citealt{matteucci2021} for a review). 

 In this paper we will focus on the Milky Way and, in particular, on the chemical evolution of the solar neighbourhood. Our main goal is to compute how many stars of different masses have contributed to build the chemical composition observed in the Solar System.
In particular, we will analyse the contributions of low- and intermediate-mass stars dying as white dwarfs (WDs), CC-SNe, and merging neutron stars (MNS).  Moreover, we will compute the number of black holes that have been created until the birth of the Solar System. 
To do that, we adopt a detailed chemical evolution model which follows the evolution of several chemical species, for a total of 43 elements from H to Pb. The adopted model derives from the two-infall model originally developed by \citet{chiappini1997} (see also \citealp{matteucci2014, romano2019}). Here, we use the revised version of  \citet{molero2023}  (see \citealp{spitoni2019,spitoni2020, spitoni2021, palla2020}), focusing our study into the solar vicinity only.

The paper is organised as follows: in Section \ref{sec_2inf}, we  will present the adopted chemical evolution model; in particular, we  will describe the prescriptions we assumed for the basic equations of chemical evolution, stellar initial mass function, star formation rate, stellar yields and  gas flows. In Section \ref{sec_results}, we will present the model results relative to the [$\alpha$/Fe] vs. [Fe/H]  trends (where $\alpha$ = O, Mg, Si, Ca). The plot of the ratio between $\alpha$-elements  ($\alpha$ = O, Mg, Si, Ca) and Fe can be used as a cosmic clock, thanks to the different timescales of production of $\alpha$s and Fe (time-delay model, \citealt{tinsley1979,matteucci2012}), and gives information on the past star formation history of the Galaxy.
In the same Section  we will provide the rates and numbers of supernovae,
white dwarfs, novae, merging neutron stars, and
black holes occurred in the solar
neighbourhood region until the formation of the Solar System.  Additionally, we  will provide the different contributions of stars to the chemical composition of the Solar System.
In Section \ref{sec_number46} we will show the results obtained for the number of stars born roughly 4.6 $\pm$ 0.1 Gyr ago with the characteristics of the Sun: this is to have an idea of how many planetary systems similar to ours might have formed  in the Galaxy. Finally, in Section \ref{conclusions} we will discuss our results and draw some conclusions.

 \section{Chemical Evolution: the Two-Infall Model}
\label{sec_2inf}
In order to discuss how different types of stars contribute to the chemical composition of the Solar System it is important to describe the original two-infall model \citep{chiappini1997}, and the revised version by \citet{molero2023} (see also \citealp{spitoni2019, spitoni2021,spitoni2024, palla2020}) that we will use in this paper.
The two-infall model suggests that the Milky Way has formed in two main gas infall events. According to the original model, the first event should have formed the   in-situ (inner) Galactic halo and the thick disc, while the second infall event should have formed the thin disc. 

 The \textit{delayed} two-infall model adopted here is a variation of the classical two-infall model of \citet{chiappini1997} developed to fit the dichotomy in the $\alpha$-element abundances observed between the thick and thin disc stars  not only in the solar vicinity (\citealp{gratton1996,fuhr1998, Hayden2014, recio-blanco2014,recioDR32022a, Mikolaitis2017})  but also at different Galactocentric distances (e.g., \citealp{hayden2015}). The model assumes that the first gas infall event formed the  thick disc whereas the second infall event, delayed by $\mathrm{\sim 3\ Gyr}$, formed the thin disc. It must be noted, that the two-infall model adopted here does not aim at distinguishing the thick and thin disc populations geometrically or kinematically (see \citealp{Kawata2016}). The first gas infall event lasts about $\mathrm{\tau_1\simeq1\ Gyr}$, while for the second event an \textit{inside-out} scenario (see e.g., \citealp{matteucci1989, romano2000, chiappini2001}) of Galaxy formation is assumed, namely, the timescale of formation by gas infall of the various regions of the thin disc increases with Galactocentric distance. 
 It should be noticed that the two main episodes described by the two-infall model are sequential in time but they are completely independent. In the original model of \citet{chiappini1997}, it was assumed a threshold gas density for star formation, which naturally produced a gap in the star formation between the end of the thick disc phase and the beginning of the thin disc and, therefore, a dichotomy in the [$\alpha$/Fe] vs. [Fe/H]  plane. However, even without the assumption of a gas threshold, the 
 double infall assumption creates a dichotomy by itself, although less pronounced, but enough to reproduce the data (see \citealp{spitoni2019}).

 \subsection{The basic equations of chemical evolution}

The basic equations which describe the evolution of the fraction of gas mass in the form of a generic chemical element \textit{i}, $G_i$, in the solar vicinity are:
\begin{equation}
    \dot{G}_i(R,t)=-\psi(R,t)X_i(R,t)+\dot{G}_{i,inf}(R,t)+\dot{E}_i(R,t),
\end{equation}
where $X_i$ is the abundance by mass of the analysed element, $\psi(t)$ is the SFR, $\dot{G}_{i,inf}(R,t)$ is the gas infall rate and $\dot{E}_i(R,t)$ is the rate of variation of the returned mass in the form of the chemical species $i$, both newly formed and restored unprocessed. This last term contains all the stellar nucleosynthesis and stellar lifetime  prescriptions.

\subsection{Star Formation Rate}

The quantity we are interested in here is the so-called stellar birthrate function, namely the number of stars with mass $dm$ which are formed in the time interval $dt$. It is factorized as the product of the SFR depending only on the time $t$, and the initial mass function (IMF), here assumed to be independent of time and being only a function of the mass $m$.

For the SFR, here we adopt as parametrization the Schmidt-Kennicutt law \citep{schmidt1959, kenni1998}, according to which the SFR is proportional to the $k$th power of the surface gas density. The SFR can then be written as:
\begin{equation}
    \psi(t) \propto \nu\sigma^{k}_{gas}(t),
\end{equation}
where $\nu$ is the efficiency of star formation, namely the SFR per unit mass of gas, and it is expressed in Gyr$^{-1}$. For the halo-thick disc phase $\nu$= 2\ Gyr$^{-1}$, whereas for the thin disc $\nu$ is a function of the Galactocentric distance $R_{GC}$, with $\nu $ (R$_{GC}$=8 kpc) $\simeq$ 1 Gyr$^{-1}$, as in \citet{molero2023} and \citet{palla2020}. It is important to highlight that gas temperature, viscosity and magnetic fields are ignored in this empirical law even if  they are expected to impact the SFRs of galaxies. Nevertheless, ignoring these parameters is a common choice for the SFR in most of galaxy evolution models.

In the scenario described by the original two-infall model there was supposed to be a gas threshold in the star formation. This created a stop in the star formation process between the formation of the thick and the thin disc. 
Here, we relax the assumption of a threshold in the gas density and the gap in the star formation is \textit{naturally} created between the formation of the two discs, since, because of the longer delay between the two infall episodes, the SFR becomes so small that a negligible number of stars is born in that time interval. 

In this context, we can make an additional distinction in the phases described by the two-infall model based on the stars that were present and dominating at each time. During the thick disc formation, the most important contribution was from core-collapse supernovae (CC-SNe) which are identified by Type II, Ib, and Ic ones, while Type Ia supernovae started giving a substantial contribution only after a time delay \citep[see][]{matteucci2021}. This important difference impacts significantly on the production of chemical elements and Galaxy composition and it is known as the \textit{time-delay model} \citep{tinsley1979,matteucci2012,matteucci2021}.

\subsection{Initial Mass Function}
The second ingredient in the stellar birthrate function is the initial mass function (IMF) which gives the distribution of stellar masses at birth and it is commonly parameterised as a power law. As to measure the IMF it is necessary to count the stars as functions of their magnitude, nowadays we can only do that for the solar region of the Milky Way (MW). We use as IMF the one proposed by \citet{kroupa1993}, which in chemical evolution is often the one which provides the best agreement with observations (see \citealt{romano2005} for a discussion). It is a three slopes IMF, with the following expression:
\begin{equation}
    \phi(m) = C 
    \begin{cases}
         m^{-(1+0.3)} & \text{if $m\leq0.5M_{\odot}$}\\
         m^{-(1+1.2)} & \text{if $0.5<m/M_{\odot}<1.0$}\\
         m^{-(1+1.7)} & \text{if $m>1.0M_{\odot}$},
    \end{cases}
\end{equation}
with $C$ being the normalization constant derived by imposing that:
\begin{equation}
1 = \int_{0.1}^{100} \! m\varphi(m) \, \mathrm{d}m,
\end{equation}
where $\varphi(m)$ is the IMF in number.

\subsection{Gas Infall}

In the case of  gas infall,
the gas is often assumed to have a primordial composition, namely with zero metal content. Since  pristine gas is enriched only in light elements such as H, He and a small part of Li and Be, the effect of the infall is that of diluting the metal content inside the Galaxy. In this work, different gas flows than the infall one (such as Galactic winds and/or Galactic fountains) are not included. In particular, Galactic fountains, which can occur in disc galaxies, have been proven not to impact in a significant manner the chemical evolution of the disc (see \citealp{melioli2009,spitoni2009}).

In the context of the delayed two-infall model \citep{molero2023} adopted here, the accretion term is computed as:
\begin{equation}
   \dot{G}_{i,inf}(R,t)=AX_{i,inf}e^{-\frac{t}{\tau_1}}+\theta(t-t_{max})BX_{i,inf}e^{\frac{t-t_{max}}{\tau_2}},
\end{equation}
where $\mathrm{X_{i,inf}}$ is the composition of the infalling gas, here assumed to be primordial for both the infall events. $\tau_1$=1 and $\tau_2$=7 Gyr are the infall timescales for the first and the second accretion event, respectively, and $t_{max}\simeq$~3.25 Gyr is the time for the maximum infall on the thin disc and it corresponds to the start of the second infall episode. The parameters $A$ and $B$ are fixed to reproduce the surface mass density of the MW disc at the present time in the solar neighbourhood. Particularly, $A$ reproduces the present time thick disc total surface mass density (12  M$_{\odot}$ pc$^{-2}$), while $B$ does the same for the present time thin disc total surface mass density (54 M$_{\odot}$ pc$^{-2}$), at the solar ring \citep{molero2023}. We remind that the $\theta$ function is the Heavyside step function.

\subsection{Element production and chemical yields}

It is worth reminding that different elements are produced in different stars. In particular:
\begin{itemize}
    \item Brown dwarfs with mass $<0.1~$M$_{\odot}$ do not ignite H, so they do not contribute to the chemical enrichment of the ISM, but they affect the chemical evolution by locking up gas;
    \item Very small stars in the mass range 0.1 M$_\odot$ - 0.8 M$_\odot$ burn only H. They die as  He-WDs on timescales longer than the age of the Universe;
    \item Low- and intermediate-mass stars (LIMS) in the mass range 0.8--8.0 M$_\odot$, contribute to the chemical enrichment through post-MS mass loss and the final ejection of a planetary nebula. They produce mainly $^{4}$He, CNO isotopes and heavy ($A>90$) s-process elements;
    \item WDs in binary systems can give rise to Type Ia SNe or novae. Type Ia SNe are responsible for producing the bulk of Fe ($\simeq$0.60 M$_\odot$ per event) and enrich the medium with tracers of elements from C to Si. They also contribute to other elements, such as C, Ne, Ca and Mg, but in a much  lower amount compared to CC-SNe. Novae can be important producers of CNO isotopes and $\mathrm{^{7}Li}$;
    \item Massive stars from 8 to 10$~$M$_{\odot}$ burn O explosively  (e-capture SNe). They produce mainly He, C and O. They leave neutron stars as remnants; 
    \item Massive stars in the mass range 10 M$_\odot$--M$_{WR}$ end their life as Type II SNe and explode by core-collapse. The explosion leads to the formation of a neutron star or a black hole, depending on the amount of mass loss during the star life and ejected material which falls back on the contracting core. M$_{WR}$ is the minimum mass for the formation of a Wolf-Rayet star. Its value depends on the stellar mass loss which in turn depends on the progenitor characteristics in term of initial mass and metallicity. For a solar chemical composition, M$_{WR}\simeq$ 25 M$_\odot$. Stars with masses above M$_{WR}$ end up as Type Ib/c and explode also by core-collapse. They are linked to the Long Gamma Ray Bursts (LGRBs) and can be particularly energetic so to be know as hypernovae  \citep[HNe,][]{paczy1998}. Massive stars are responsible for the production of most of $\alpha$-elements (such as O, Ne, Mg, Si, S, Ca), some Fe-peak elements, light (A<90) s-process elements (especially if stellar rotation is included) and may contribute also to r-process nucleosynthesis (if strong magnetic field and fast rotation are included).
    \item  Mergers of compact object and in particular neutron star and black hole binary systems can be  powerful sources of r-process material.
\end{itemize}

The stellar yields that we adopt for stars of all masses, Type Ia SNe and merging neutron stars are similar to those adopted in \citet{romano2010} and \citet{molero2023}. In particular, for massive stars we adopt the yields of \citet{koba2006} and the Geneva group \citep{meynet2002,hirschi2005,hirschi2007,ekstrom2008} for what concerns the CNO elements. For LIMS yields we assume those of \citet{karakas2010}; for Type Ia SNe, those of \citet{iwamoto1999}  and for neutron capture elements those adopted by \citet{molero2023} for merging neutron stars as well as for massive stars dying as magneto-rotational supernovae.

\section{Results}

\label{sec_results}
\subsection{The star formation rate}
\begin{figure}
    \centering
    \includegraphics[width=\columnwidth]{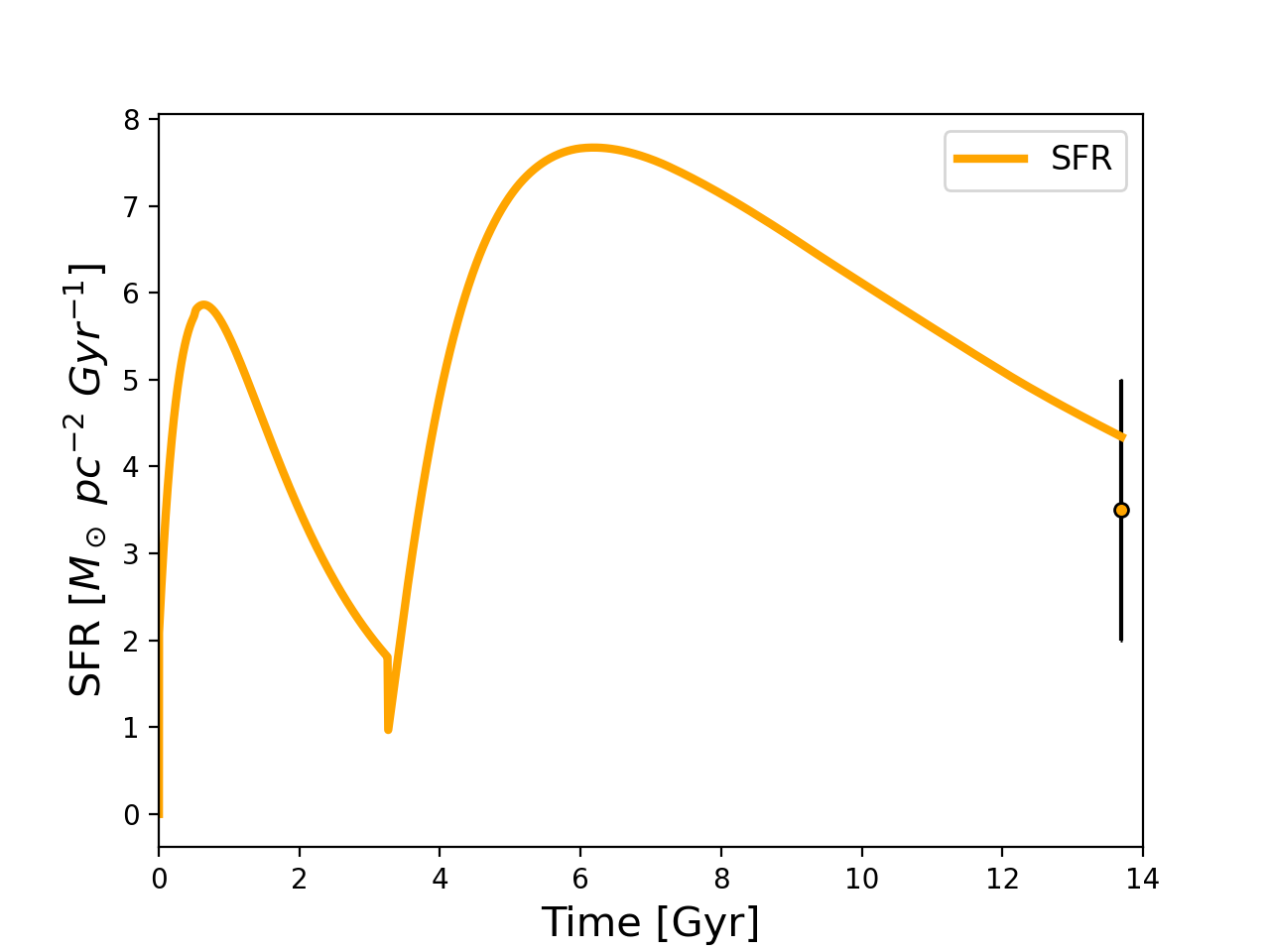} 
    \caption{The evolution with time of the SFR, as predicted by the two-infall model for the solar vicinity. Observed present-day value is from \citet{prantzos2018}.}
    \label{SFR}
\end{figure}

Before presenting the analysis of the abundance patterns, it is important to compare the evolution of the SFR predicted by our model at R$_{\text{GC}}$=8 kpc to present-day observations in the solar vicinity. The SFR, expressed in units of M$_\odot$ pc$^{-2}$ Gyr$^{-1}$, is shown in Figure \ref{SFR}.  The gap between the two different disc phases, as discussed before, is clearly visible and the present-day value predicted by our model appears to be in nice agreement with the measured value in the solar neighborhood, as suggested by \citet{prantzos2018}.

To compute how many solar masses of stars have been formed until the moment of the formation of the Solar System, we computed the integral of the SFR in the time interval 0.0 - 9.2 Gyr, as:
\begin{equation}
    \int_{0.0}^{9.2 \, \text{Gyr}} \psi(t) \,dt= 51\ M_{\odot}pc^{-2}.
\end{equation}
This value, once multiplied by the area of the solar annular ring 2 kpc wide ($\sim$ 10$^8$ pc$^2$), gives the total mass of stars ever formed, equal to 5.1 $\times$ 10$^9$ M$_\odot$. We stress that this quantity takes into account also the contribution from the stellar remnants (namely white dwarfs, neutron stars and black holes).

For what concerns the total metallicity in the ISM 4.6 Gyr ago, we predict Z$_{\odot}$=0.0130, in excellent agreement with the solar metallicity by \citet[][Z$_{\odot}$=0.0134]{asplund2009}, and the predicted Fe abundance is 12 + log(Fe/H)$_{\odot}$=7.48, again in excellent agreement with the observed abundance.

\subsection{Analysis of the [$\alpha$/Fe] vs [Fe/H] plot by means of the time-delay model}

\begin{figure}[h!]
    \centering
    \includegraphics[scale=0.62]{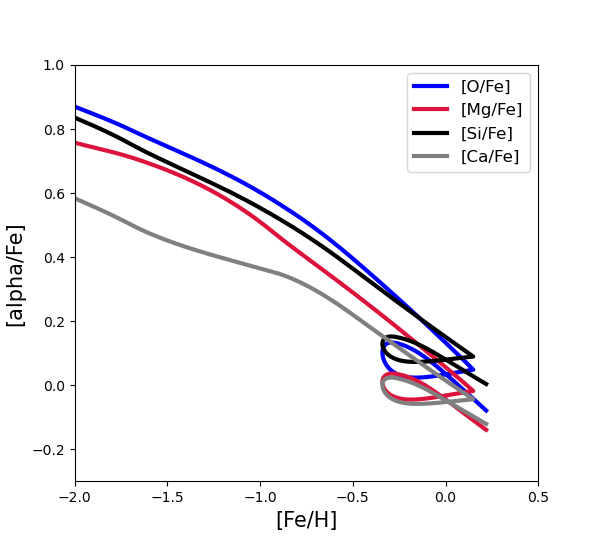} 
    \caption{ [$\alpha$/Fe] versus [Fe/H] abundance ratios predicted  by our fiducial chemical evolution model for different $\alpha$-elements: oxygen (blue line), magnesium (red line), silicon (black line) and calcium (grey line), for the solar vicinity.}
    \label{alpha_fig}
\end{figure}

In this section, we will present and analyse the [$\alpha$/Fe] vs. [Fe/H] abundance patterns  \footnote{We remind that the notation [X/Y] has the meaning [X/Y]= log(X/Y)$-$log(X/Y)$_\odot$ with X (Y) being the abundance by number of the element X (Y).}  of some $\alpha$-elements, namely O, Mg, Si and Ca.
Figure \ref{alpha_fig} shows the plots of [$\alpha$/Fe] vs. [Fe/H] ($\alpha$ =O, Mg, Si, Ca) as predicted by the model. The time-delay model \citep{tinsley1979,matteucci2012} provides a satisfying explanation for these paths: the ratio of [$\alpha$/Fe] at very low metallicity is rather flat  and (the slope of the `flat' portion is due to the different nucleosynthetic yields of different $\alpha$-elements). 

Because only CC-SNe produce $\alpha$ elements in a substantial way plus some amount of Fe, the flat part of the plot is representative only of the contribution to the [$\alpha$/Fe] ratio from massive stars at early times. When [Fe/H] $\geq -$1.0 dex, Type Ia SNe start giving their contribution, as it can be seen from the change of the slope shown in the plots. This happens because, as it was already explained, Type Ia SNe are the main producers of Fe and they eject this element into the ISM on longer timescales. The loop shown by the curves in  Figure \ref{alpha_fig}, is due to the gap in the star formation occurring in between the two infall events. In fact, as explained in \citet{spitoni2019}, the second infall causes a dilution of the absolute abundances, producing a horizontal behavior in the [Fe/H] at almost constant [$\alpha$/Fe]. Then, when the star formation recovers, the [$\alpha$/Fe] ratio rises and then decreases slowly again because of the advent of Type Ia SNe. These loops can successfully explain the bimodality in [$\alpha$/Fe] ratios \citep{spitoni2019,spitoni2020}.

If the X-axis can in principle be interpreted as a time axis, the [$\alpha$/Fe] vs. [Fe/H] relation can be used to extract the timescale for the formation of the thick and thin discs, knowing that thick disc stars have metallicity $\le -$0.6 dex. Originally, \citet{matteucci1986} derived the timescale of the formation of the inner halo-thick disc to be around 1.0-1.5 Gyr. Subsequent studies dealing with detailed evolution of the thick disc have confirmed a timescale of $\sim$1 Gyr for its formation \citep[e.g.][]{micali2013,grisoni2017}. Here, we find the same timescale. It is worth noting that the timescale of formation of the thin disc at the solar ring is provided by the fit to the G-dwarf metallicity distribution and is $\sim$ 7 Gyr \citep[e.g.][]{chiappini1997,grisoni2017}.

\subsection{Rates and numbers of  supernovae, white dwarfs, novae, merging neutron stars, and black holes} 
\label{sec_rates}
Since we reproduce quite well the solar metallicity and the abundance patterns in the solar vicinity, now we can proceed to compute in detail the rates and numbers of supernovae (Type Ia and core-collapse), white dwarfs, novae, neutron stars and black holes  occurred until the formation of the Solar System.  
  We define, unless otherwise stated,  as solar vicinity the annular region centered in the Sun, as defined before, while for the whole disc we assume an area of approximately 10$^9$ pc$^2$.

\subsubsection{Type Ia Supernovae}

 To compute the number of Type Ia SNe exploded until the formation of the Solar System, we proceed just in the same way as for the total number of stars formed (see previous Section). In particular, we compute their rate as the fraction of WDs in binary systems that have the necessary conditions to give rise to a Type Ia SN event. This allows us to compute the rate of Type Ia SNe as suggested by \citet{greggio2005}:
\begin{equation}
    (Rate)_{SNeIa}(t)= K_{\alpha} \int_{\tau_i}^{min(t,\tau_x)} A(t-\tau) \psi(t-\tau) DTD(\tau) \, d\tau,
\end{equation}
where $\tau$ is the total delay time, namely the nuclear stellar lifetime of the secondary component of the binary system plus a possible delay due to the gravitational time delay in the DD model. $A(t-\tau)$ is the fraction of binary systems which give rise to SNe Type Ia and we assume it constant in time. The DTD($\tau$) is the Delay Time Distribution function, describing the rate of explosion of Type Ia SNe for a single starburst. The DTD is normalised as:
 \begin{equation}
 \int_{\tau_i}^{\tau_x} DTD(\tau) \, d\tau=1,
 \end{equation}
with $\tau_{i}$ being the lifetime of a $\sim$8$~$M$_{\odot}$ star and $\tau_{x}$ the maximum time for the explosion of a Type Ia SN. Here, we adopt the DTD for the wide DD scenario as suggested by \citet{greggio2005}, where a detailed description can be found (see also \citealp{simonetti2019, molero2021td}). Finally, $K_{\alpha}$ is a function of the IMF, namely:
\begin{equation}
K_{\alpha}= \int_{0.1~M_{\odot}}^{100~M_{\odot}} \varphi(m) \, dm.
\end{equation}
The predicted present-time Type Ia SN rate  for the whole disc is: 
\begin{equation}
(Rate)_{SNeIa, \, current}=0.40 \cdot events/century.
\end{equation}
It is important to notice that this result is in agreement with the observed rate of 0.43 events/century  \citep{cappellaro1997,li2010}  which confirms the validity of the model.

We then compute the number of Type Ia SNe until the birth of the Solar System, in the solar vicinity. To do so, we integrate the rate from 0$~$Gyr to 9.2$~$Gyr, obtaining: 
\begin{equation}
   N_{SNeIa}(t)= \int_{0}^{9.2 \, Gyr}  (Rate)_{SNeIa}(t) \,dt= 2.87 \cdot 10^{6}.
\end{equation}

\subsubsection{Core-Collapse Supernovae}

We compute the fraction of massive stars that will die as CC-SNe by assuming that they originate from single massive stars or massive binaries. The rate of Type II supernovae is computed as:
\begin{equation}
    (Rate)_{SNeII}(t)= \int_{8~M_{\odot}}^{M_{WR}} \psi(t-\tau_{m})\varphi(m) \,dm,
\end{equation}
where, as previously described, M$_{\text{WR}}$ is the limiting mass for the formation of a Wolf-Rayet star. 
The rate of Type Ib/Ic SNe can be calculated as \citep[see][]{bissaldi2007}:

\begin{equation}
\begin{split}
  (Rate)_{SNeIb,c}(t) & = (1-\gamma) \int_{M_{WR}}^{M_{max}} \psi(t-\tau_m) \phi(m) \,dm \\
& + \gamma \int_{14.8~M_{\odot}}^{45~M_{\odot}} \psi(t-\tau_m) \phi(m) \,dm,
\end{split}
\end{equation}
\noindent where the parameter $\gamma$ is chosen to reproduce the number of massive binary systems in the range 14.8 $\div$ 45$~$M$_{\odot}$ as proposed by \citet{yoon2010}  to produce a SNeIb,c. The mass M$_{max}$ is the maximum mass allowed by the IMF, equal to 100 M$_{\odot}$.

Considering both SNeII and SNeIb,c, we obtain a total rate of CC-SNe of $(Rate)_{CCSNe, \, current}$=2.23$~$events/century which is in agreement with the observational data of 1.93$~$events/century  \citep{cappellaro1997}, as shown in Table  \ref{tab_SN}.
The total number of CC-SNe exploded until the formation of the Solar System, in the solar vicinity, is:
\begin{equation}
    N_{CCSNe}(t) = \int_{0}^{9.2 \, Gyr} (Rate)_{CCSNe}(t)  \, dt =   26.47 \cdot 10^6. 
\end{equation}

\begin{figure}
   \centering
   \includegraphics[scale=0.6]{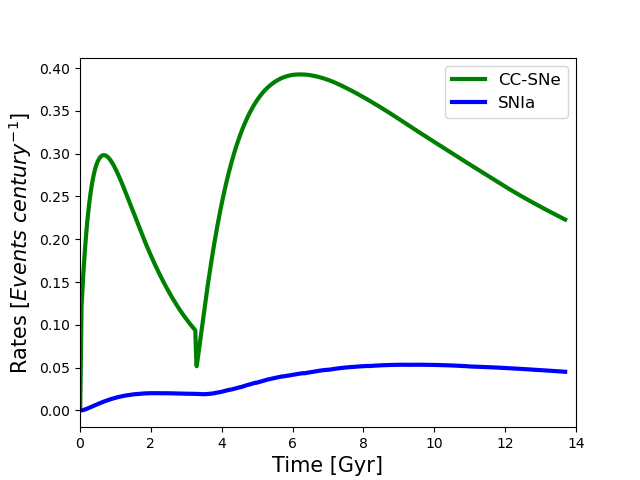} 
    \caption{Predicted rate of core-collapse  (green line) compared to that of Type Ia supernovae (blue line) in the solar vicinity.}
   \label{fig:CC}
\end{figure}

 In Figure \ref{fig:CC} we can see the CC-SN rate behaviour and appreciate the fact that, as expected, it follows the SFR path, namely it shows the same gap as the SFR versus time. In fact, all the quantities related to the SFR, such as CC-SN rate, formation of neutron stars and black hole rates show a dip corresponding to the strong decrease in the star formation occurring between the formation of the thick and thin disc.  In the same Figure we show the Type Ia SN rate as a function of time, in the solar vicinity. 
 In Table \ref{tab_SN}, we finally summarise our results compared to observational data. It is worth noting that the observed current rates of SNe, as well as those of novae and MNS, are derived for the entire MW, while we show in the figures the predicted rates for the solar ring. The predictions shown in Table \ref{tab_SN}, as well as in the other Tables, refer instead to the entire disc, and,
as it can be seen, the agreement between our predictions and data is quite good. Concerning the observed current rates for the solar vicinity, we could perhaps rescale those for the entire disc to the area of the solar vicinity. This would simply mean dividing the disc rates by a factor of ten.

\begin{table*}
\tiny
 \begin{center}
 \caption{Comparison between observational data relative to SN rates \citep{cappellaro1997} compared with model results. We can see that all theoretical values are consistent with observational data. In the first column there are the observed rates, in the second column the predicted ones and in the third column are the computed total numbers of SNe exploded from the beginning up to the formation of the Solar System.}
 \label{tab_SN}
   \begin{tabular}{|c|c|c|c|}
   \hline
   & \textbf{Observational Data}  & \textbf{Rates} & \textbf{Numbers}  \\
     & (whole disc, present-day)  & (whole disc, present-day) & (solar vicinity, 9.2 Gyr of evolution)   \\
   [2ex] \hline 
   \textbf{SNeIa} & $0.43~SNe/century$ & $0.45~SNe/century$ & $2.87~million$ \\ [2ex] 
   \hline
   \textbf{CC-SNe} & $1.93~SNe/century$ & $2.23~SNe/century$ & $26.47~million$ \\[2ex] \hline
   \end{tabular} 
    \end{center}
\end{table*}

\subsubsection{White Dwarfs and Novae}

\begin{figure}
    \centering
    \includegraphics[scale=0.6]{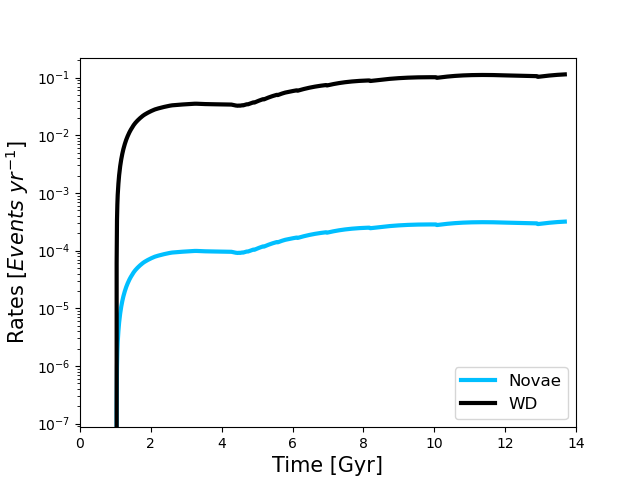} 
    \caption{Comparison of  the rate of formation of nova systems and white dwarfs, in the solar vicinity.}
    \label{fig:wd}
\end{figure}

In Figure  \ref{fig:wd}, we plot  the rate of formation of WDs, originating in the mass ranges 0.8-8 M$_{\odot}$, from which we obtain the number of white dwarfs that formed until the moment of formation of the Solar System, in the solar vicinity. This is computed as: 
\begin{equation}
   N_{WD}(t) = \int_{0}^{9.2 \, Gyr}  (Rate)_{WD}(t) \,dt= 423.88 \cdot 10^6 .
\end{equation}

A nova outburst is caused by the thermonuclear runaway on top of a white dwarf accreting H-rich matter from a close companion (a main sequence or a giant star) that overfills its Roche lobe. The system survives the explosion and the cycle is repeated some 10$^4$ times.

We compute the nova rate by assuming that it is a fraction of the white dwarf rate. To do so, it is appropriate to define a parameter, $\alpha_{nova} < 1$, that represents the fraction of white dwarfs that will form novae, and it is tuned to reproduce the present time nova rate in the Galaxy. In this work, the value that we used was $\alpha_{nova}$=0.0028 and that allowed us to correctly reproduce the observed nova rate in the Galaxy, which is 20 $\div$ 40$~$events/yr \citep{dellavalle2020}. Indeed, our model prediction is $(Rate)_{Novae, \, current}=31~$ number/yr.
There are various ways to compute the rate of novae in our Galaxy, such as using the known novae to extrapolate for those too far to be seen or else observing novae in another galaxy and extrapolate their rate in the Milky Way by assuming that every nova from the other galaxy can be seen.

In particular, we define the rate of novae as:
\begin{equation}
    (Rate)_{Novae}(t)=\alpha_{nova} \int_{0.8~M_{\odot}}^{8~M_{\odot}} \psi(t- \tau_{m_2}-\Delta t) \varphi(m) \,dm,
\end{equation}
with $\Delta t$ being the delay time between the formation of the WD and the first nova outburst (the WD needs to cool down before the nova outburst can occur) and $\tau_{m_2}$ the lifetime of the secondary star that determines the start of the mass accretion onto the WD.

 We had to consider that every nova system produces $10^4$ nova outbursts, so if we want to compute the nova rate we need to multiply  the nova formation rate by this number. In Figure \ref{fig:wd}, we show the rates of WDs and novae together as  functions of time. In Table \ref{tab_NOVA} we report the total rates and total numbers of WDs, novae and nova outbursts.

The number of nova systems and nova outbursts that occurred until the moment of the formation of the Solar System, in the solar vicinity, is computed as:

\begin{equation}
    N_{Novae}(t) = \int_{0}^{9.2 \, Gyr}  (Rate)_{Novae} (t) \,dt= 1.18\cdot 10^{6},
\end{equation}
that lead us to the following result for the number of nova outbursts:

\begin{equation}
    N_{NO}(t) = \int_{0}^{9.2 \, Gyr}  (Rate)_{NO}(t) \,dt= 1.18\cdot 10^{10}.
\end{equation}

The numbers and rates of WDs and novae are presented in Table2.

\begin{table*}
\tiny
 \begin{center}
 \caption{Comparison between observational data about nova outbursts \citep{dellavalle2020}  and model results. {\bf  Theoretical rates of formation of WDs and nova systems are also reported, though observed rates of formation of WDs and nova systems are not available.} The legenda is like in Table 1.}
   \label{tab_NOVA}
   \begin{tabular}{|c|c|c|c|}
   \hline
   & \textbf{Observational Data}  & \textbf{Rates} & \textbf{Numbers}  \\
     & (whole disc, present-day)  & (whole disc, present-day) & (solar vicinity, 9.2 Gyr of evolution)   \\
   [2ex] \hline 
   \textbf{Nova systems} & $-$ & $0.0031~events/yr$ & $1.18~million$ \\ [2ex] \hline
   \textbf{Nova outbursts} & $25-30~events/yr$ & $31~events/yr$ & $11.8~billion$ \\[2ex] \hline
   \textbf{White dwarfs} & $-$ & $1.11~events/yr$ & $423~million$ \\[2ex] \hline
   \end{tabular} 
   \end{center}
\end{table*}

\subsubsection{Neutron Stars}
Neutron stars are among the densest objects known with an average density of around 10$^{14}~$g/cm$^3$. They are remnants of massive stars, but it is not clear the upper mass limit for the formation of a neutron star. In fact, if the stellar core is larger than the so-called Oppenheimer-Volkoff mass ($\sim$ 2 M$_{\odot}$), then a black hole will form. The limiting initial stellar mass between the formation of a neutron star and a black hole is strongly dependent upon assumptions in stellar models, such as for example the rate of mass loss during the evolution of massive stars. 
In the model adopted here to compute the rate of neutron stars, we assumed that stars with masses from 9 to 50 M$_{\odot}$ \citep{molero2023} leave a neutron star after their death.  With this assumption we found that the current rate of formation of neutron stars is $(Rate)_{NS, \, current}\simeq 2.93\times10^{4}$ number/Myr.

\subsubsection{Merging Neutron Stars}
\begin{figure}[ht!]
    \centering
    \includegraphics[scale=0.6]{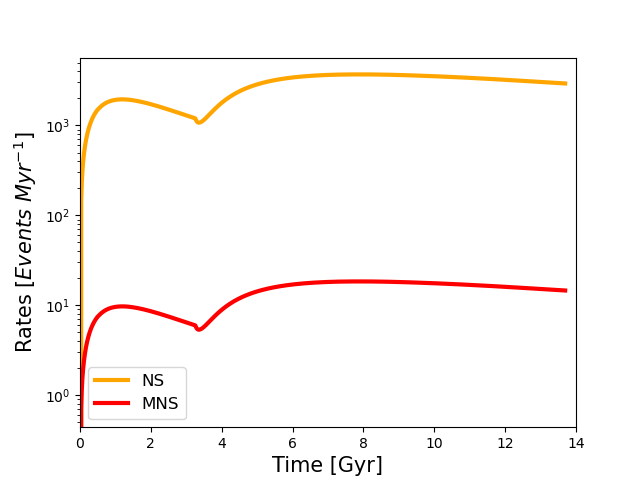} 
    \caption{Rates of neutron star (NS, yellow line) formation  and  merging neutron star (MNS, red line) predicted by our chemical evolution model for the solar vicinity.}
    \label{fig:NS}
\end{figure}

Merging Neutron Stars (MNS) are important for what concerns the chemical evolution of galaxies, as they produce r-process elements. It was confirmed by the gravitational event GW170717 \citep{abbott2017}  that the merging of neutron stars can produce a strong gravitational wave and that their contribution to the chemical composition of galaxies cannot be ignored. The rate of MNS and their number, is assumed to be proportional to the rate of formation of neutron stars (as proposed by \citealp{matteucci2014}), namely:
\begin{equation}
(Rate)_{MNS}(t)= \alpha_{NS} \cdot (Rate)_{NS}.
\end{equation}
The constant $\alpha_{NS}$ is set to $\mathrm{\sim10^{-3}}$, chosen to correctly reproduce the observational rate of $83^{+209.1}_{-66.1}$ MNS$/Myr$ \citep{kalogera2004} in the Milky Way.
Finally, the total number of neutron stars and MNS that contributed to the chemical composition of the Solar System, in the solar vicinity, were obtained as the time integral of their rates, namely:
\begin{equation}
   N_{NS}(t) = \int_{0}^{9.2 \, Gyr}  (Rate)_{NS} (t)\,dt= 23.15 \cdot 10^{6},
\end{equation}

\noindent and
\begin{equation}
   N_{MNS}(t) = \int_{0}^{9.2 \, Gyr}  (Rate)_{MNS} (t) \,dt= 0.11 \cdot 10^{6}.
\end{equation}
 A plot with both neutron stars and MNS rates is provided in Figure \ref{fig:NS}. The numbers  and rates of neutron stars and MNSs can be found in Table \ref{tab_NS}.

\begin{table*}[h!]
\tiny
 \begin{center}
  \caption{Comparison between observational data relative to the MNS rate in the Milky Way \citep{kalogera2004} and model results for neutron star formation and MNS rate. We can see that the computed value for the MNS rate is consistent with observational data and results by \citet{molero2021td}. Notice that these values were obtained with $\alpha_{NS}=0.005$}
  \label{tab_NS}
   \begin{tabular}{|c|c|c|c|}
   \hline
   & \textbf{observational Data}  & \textbf{Rates} & \textbf{Numbers} \\
     & (whole disc, present-day)  & (whole disc, present-day) & (solar vicinity, 9.2 Gyr of evolution)   \\[2ex] \hline 
   \textbf{Neutron Stars} & $-$ & $29261~events/Myr$  & $23.15~million$ \\ [2ex] \hline
   \textbf{MNS} & $83^{+209.1}_{-66.1}~events/Myr$ & $146~events/Myr$ & $0.11~million$ \\[2ex] \hline
   \end{tabular} 
  \end{center}
\end{table*}

\begin{figure}
    \centering
    \includegraphics[scale=0.6]{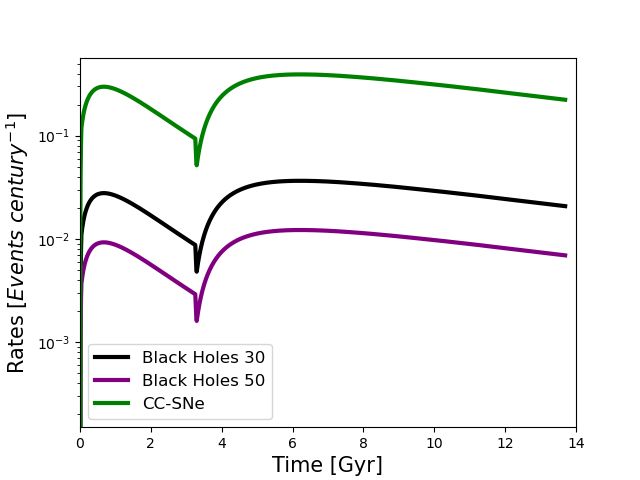} 
    \caption{Comparison between the rate of black holes that come from stars with masses $\geq$ 30$~$ M$_{\odot}$ (black line) and from stars with masses $\geq$ 50$~$ M$_{\odot}$ (purple line) with  CC-SNe (green line), in the solar vicinity.}
    \label{fig:BH}
\end{figure}

\begin{table}[h!]
\tiny
 \begin{center}
   \caption{Comparison between different model results about the number of black holes with M$_{BH}\geq$ 30$~$ M$_{\odot}$ and M$_{BH} \geq$ 50$~$M$_{\odot}$ as a percentage of massive stars.}
   \label{tab_BH}
   \begin{tabular}{|c|c|c|}
   \hline
   \textbf{Black holes} & \textbf{$M_{BH} \geq 30~M_{\odot}$}  & \textbf{$M_{BH} \geq 50~M_{\odot}$}  \\[2ex] \hline 
   \textbf{Percentage} & $9.14\%$ & $3.00\%$\\ [2ex] \hline
   \textbf{Number} & 2.46 $million$ & 0.82 $million$ \\ [2ex] \hline
   \end{tabular} 
 \end{center}
\end{table}

\begin{figure}
    \centering
    \includegraphics[scale=0.43]{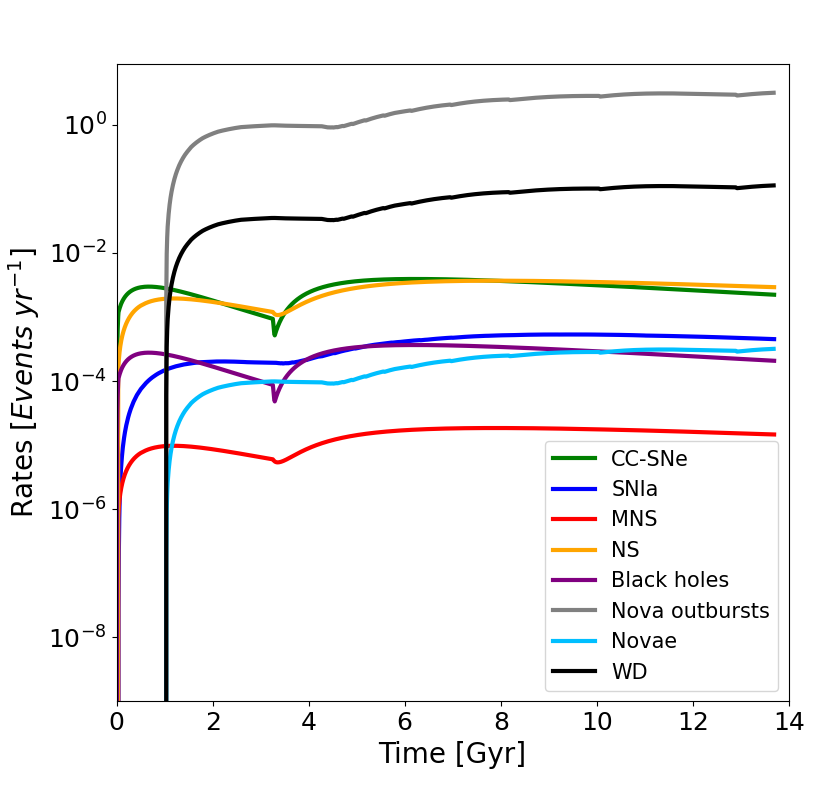} 
    \caption{Plot of the rates of all types stars in the solar vicinity, as discussed up until now. Notice that some types of stars, namely white dwarfs, novae and nova outbursts, started forming after t $\sim$ 1 Gyr and not from the Big Bang.}
    \label{fig:ALL}
\end{figure}

\begin{figure*}
    \centering
    \includegraphics[scale=0.35]{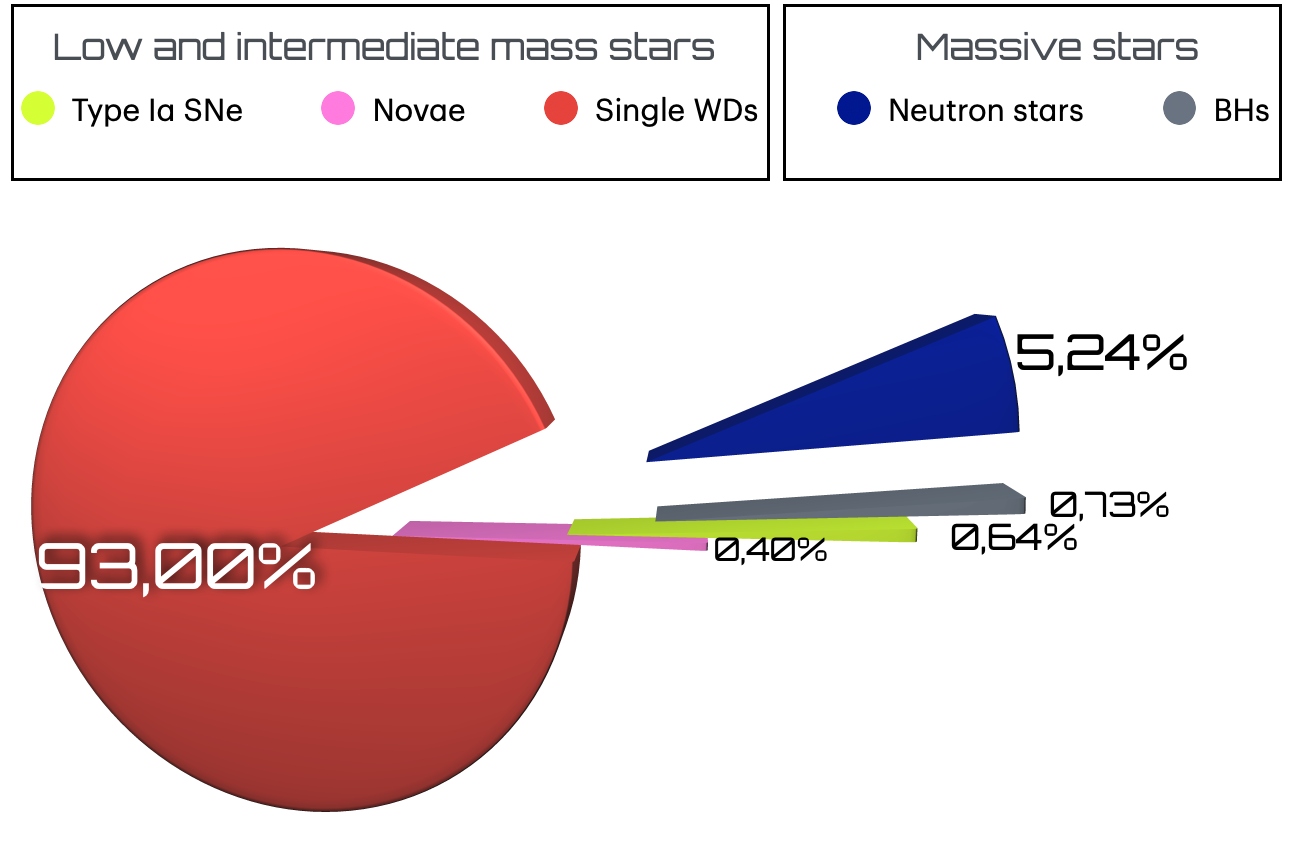} 
    \caption{Percentage contributions of the stars analysed in this paper (identified with their final outcomes) to the chemical composition of the Solar System.}
    \label{pie}
\end{figure*}

\subsubsection{Black Holes}

The last rate that we computed is the rate of birth of black holes originating from the massive stars that can leave a black hole after their death.  The rate of formation of black holes is:
\begin{equation}
    (Rate)_{BH}(t)= \int_{M_{BH}}^{100~M_{\odot}} \psi(t)\varphi(m) \,dm
\end{equation}

In our model, we assumed two different values of M$_{BH}$ (the limiting initial stellar mass for having a black hole as a remnant), namely M$_{BH}$=30$~$M$_{\odot}$ and M$_{BH}$=50$~$M$_{\odot}$. The total number of black holes that formed until the formation of the Solar System is computed as:
\begin{equation}
   N_{BH} = \int_{0}^{9.2 \, Gyr} (Rate)_{BH}(t) \, dt .
\end{equation}
The first choice, M$_{BH}$=30$~$M$_{\odot}$, led us to the result that roughly 9.14$\%$ of massive stars will leave a black hole, which means N$_{BH} \sim$ 2.46 $\cdot$ 10$^6$, while for M$_{BH}$=50$~$M$_{\odot}$ the number drops to 3.00$\%$, that is N$_{BH} \sim$ 0.82 $\cdot$ 10$^6$.

In Figure \ref{fig:BH}, we provide a comparison between the rate of black holes under the assumptions of M$_{BH} \geq$ 30$~$M$_{\odot}$ and M$_{BH} \geq$ 50$~$M$_{\odot}$, as well as the rate of CC-SNe. 
 
\subsubsection{Comparison of all the rates}

To have a complete vision of all types of stars that contributed to the formation of the Solar System and to its chemical composition it is interesting to plot the different rates together, so that it is possible to better compare them.

\begin{figure*}
    \centering
 \includegraphics[scale=0.33]{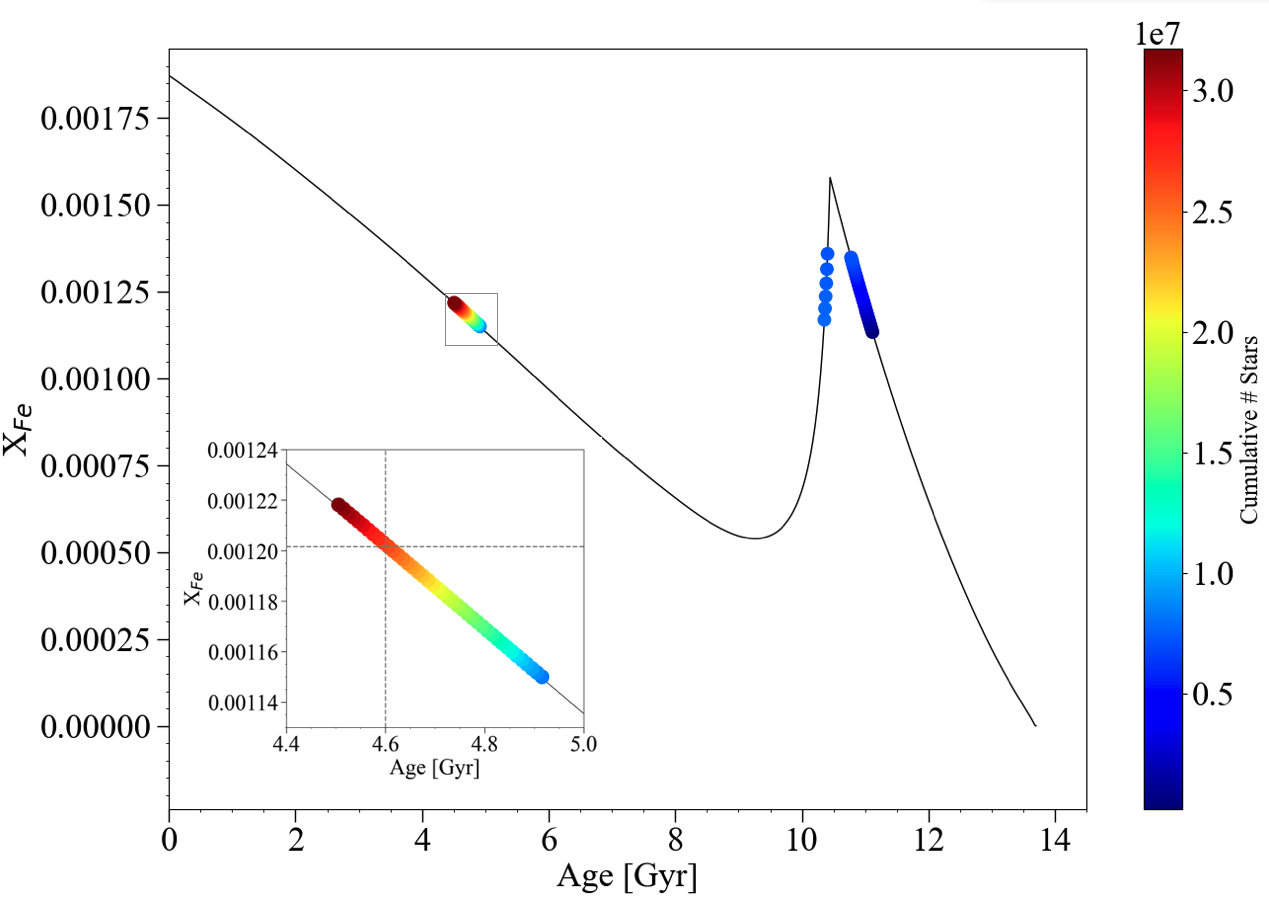} 
    \caption{ Iron mass fraction X$_{Fe}$ versus Galactic age, obtained with our chemical evolution model for the solar vicinity. With the colour-coded points, we highlight the cumulative number of stars sharing the same physical and chemical properties of our Sun formed from the beginning up to 4.6  $\pm$ 0.1 Gyr ago (see Section \ref{sec_number46} for further details). In the inset plot we zoom-in the region with predicted Sun-like stars  younger than  4.91 Gyr. The horizontal dashed line  indicates the iron mass fraction predicted for our Sun born 4.6 Gyr ago (vertical dashed line).} 
    \label{fig_ALL_SUNS}
\end{figure*}

In particular, in Figure \ref{fig:ALL}, we report all the rates discussed up to now together for comparison. It is clear from the Figure that the nova outbursts, for the assumptions made, represent the largest number of events, but the material ejected during each burst is much less than what is produced by SNe and MNS. However, novae cannot be neglected in chemical evolution models since they can be responsible for the production of some important species.  We underline that black holes (M$_{BH}$=30$~$M$_{\odot}$)  are represented in this graph even if they do not contribute to the chemical enrichment, but they are related to very massive stars that can eject large amounts of metals before dying. Moreover, stars leaving black holes as remnants  (Type Ib and Ic SNe) seem to be related to long GRBs, and the rate of formation of black holes can therefore trace the rate of 
these events (see \citealt{bissaldi2007}).

In Figure \ref{pie}, we show a stellar pie illustrating the different percentages of stellar contributors to the chemical composition of the Solar System.
Clearly, the majority of stars ever born and dead from the beginning to the formation of the Solar System are belonging to the range of low and intermediate masses; these stars have mainly contributed to the production of He, some C, N and heavy s-process elements, while the massive stars, whose remnants are neutron stars and black holes, have produced the bulk of $\alpha$-elements, in particular O which dominates the total solar metallicity Z. On the other hand, the bulk of Fe originated from Type Ia SNe. The novae can be important producers of CNO isotopes \citep[see][]{romano2022} and perhaps $^{7}$Li 
\citep[see][]{izzo2015,cescutti2019,matteucci2021_li}. Concerning r-process elements, the most reasonable assumption is that they have been produced in the range of massive stars both from merging neutron stars and, eventually, some peculiar type of CC-SNe \citep[see][]{simonetti2019,molero2023}.

\section{The number of stars similar to Sun born from the beginning up to the formation of the Solar System}
\label{sec_number46}

In order to investigate a crucial aspect of   the  general argument  of the  evolution of the Galactic population of stars and their habitable planets,  let us introduce the concept of the number of  solar twins  born before our Sun.       
Although it is known that also M stars (0.08-0.45 M$_{\odot}$) can host Earth-like planets and the numbers of these stars have been computed for the Milky Way (see \citealt{spitoni2017}),  here we focus only on twins (life bearing) of our Sun, as hosts of Earth-like planets.
Hence, we compute the number of stars in the range of  mass 0.92-1.08 $M_{\odot}$ born from the beginning up to 4.6 $\pm 0.1$ Gyr ago and with solar Fe abundance compatible (within 1$\sigma$) with the value from \citet{asplund2009} who reported 7.50 $\pm$ 0.04 dex.
  In other words, this quantity, $N_\odot$, is the number of "twins" formed in the vicinity of the current location of the Sun until 4.6 Gyr ago. Such a time limit is set because we take the working assumption that, on average, intelligent life develops in the twin solar systems  within the same time required on Earth. We then obtain: $N_\odot \simeq 31.70 \times 10^6$.
Among $N_{\odot}$, our Sun is the 
$\sim$ 2.61 $\cdot$  10$^7$-th star born in solar vicinity with a predicted Fe abundance of 7.48 dex, in excellent agreement with the observed one.

In Figure \ref{fig_ALL_SUNS}, we show the Fe abundance by mass as a function of Galactic age, as predicted for the solar vicinity. The peak of the Fe abundance at early times corresponds to the formation of the thick disc; then, it follows a gap due to the strong depression of the SFR between the formation of the thick and thin discs, and finally an increase of the Fe abundance up to the time of formation of the Solar System and beyond. 
Additionally, the figure shows the cumulative number of stars with the same mass and Fe abundance formed up to the appearance of our Sun. Notably, at early times, during the age interval from 11.12 Gyr to 10.36 Gyr ago, a total of  $\sim$ 0.77 $\times$ 10$^{7}$ solar-like stars were already formed. Then, due to the strong metallicity dilution by the  infalling gas with a pristine chemical composition associated with the formation of the thin disc, the Fe abundance decreased and remained sub-solar until 4.91 Gyr ago. In more recent times, i.e. Galactic ages between 4.91 and 4.50 Gyr,  $\sim$ 2.40 $\times$ 10$^{7}$  Suns were  formed ($\sim$ 75.6$\%$ of the total number).

\section{Conclusions and discussion}
\label{conclusions}
In this work  we have calculated the rates and the relative numbers of stars of different masses, which died either quiescently or in an explosive way as SNe, that contributed to the chemical composition of the Solar System (which formed about 4.6 Gyr ago) in the context of the two-infall model for the chemical evolution of the Milky Way.\\
Our results for each type of stars residing in the solar vicinity, can be summarised as follows:
\begin{itemize}
    
    \item \textbf{Number of Type Ia supernovae}: $2.87~millions$
    \item \textbf{Number of core-collapse supernovae}: $26.47~millions$
    \item \textbf{Number of white dwarfs}: $423.88~millions$
    \item \textbf{Number of nova systems}: $1.8~millions$
    \item \textbf{Number of nova outbursts}: $1.8 \cdot 10^4~millions$
    \item \textbf{Number of neutron stars}: $23.5~millions$
    \item \textbf{Number of merging neutron stars}: $0.11~millions$
    \item \textbf{Number of black holes ($M \geq 30~M_{\odot}$) }: $2.46~millions$
    \item \textbf{Number of black holes ($M \geq 50~M_{\odot}$) }: $0.82~millions$
   \item\textbf{Number of solar twins born 4.6 Gyr ago}: $31~millions$
   \item\textbf{Number of stars ever born and still alive 4.6 Gyr ago}: $35 ~ hundreds \, of \, millions$.
    \end{itemize}
 It is worth noting that all these numbers should be divided by 25 if one wants to restrict the solar vicinity area to a square centered in the Sun with a side of 2 kpc.  
 Concerning the percentage of black holes that formed until the birth of the Solar System, in relation to the number of massive stars (8$~$M$_{\odot} \leq$ M $\leq$ 100$~$M$_{\odot}$), we found that only 3$\%$ of massive stars have the necessary characteristics to become black holes if we assume a limiting mass for the formation of black holes  $\ge$50 M$_{\odot}$, while the percentage increases to 9\% if we accept stars with M $\geq$ 30$~$M$_{\odot}$. 

 Finally, since our aim was to understand how different types of stars contributed to the chemical composition of the solar neighbourhood, in particular the metallicity, we have estimated the percentage of the contributions of different stars to the $\alpha-$elements and Fe which are the major constituents  of the metallicity $Z$. In particular: $^{16}$O is almost entirely  (98.5\%) produced by stars with masses $>8M_{\odot}$, so is $^{24}$Mg (98.2\%). For the other $\alpha$-elements, Ca and Si, we find that they are produced by 74.5\% and 79 \%, respectively, in massive stars, and the rest by Type Ia SNe. Finally, Fe is produced by 70\% by Type Ia SNe and by 30\% by massive stars, a result already known (\citet{matteucci1986}). It is worth noting that these percentages depend on the assumed stellar yields and IMF. However, since both yields and IMF are suitable for the solar vicinity we are confident that the percentages are correct.

\section*{Acknowledgement}

 F. Matteucci, M. Molero and A. Vasini thank I.N.A.F. for the 1.05.12.06.05 Theory Grant - Galactic archaeology with radioactive and stable nuclei. F.~Matteucci also thanks Ken Croswell for stimulating the computation of the total number of novae, thus giving the idea for the present paper. This research was supported by the Munich Institute for Astro-, Particle and BioPhysics (MIAPbP) which is funded by the Deutsche Forschungsgemeinschaft (DFG, German Research Foundation) under Germany´s Excellence Strategy – EXC-2094 – 390783311. F. Matteucci thanks also support from Project PRIN MUR 2022 (code 2022ARWP9C) "Early Formation and Evolution of Bulge and HalO (EFEBHO)" (PI: M. Marconi), funded by the European Union – Next Generation EU. E. Spitoni thanks I.N.A.F. for the  
1.05.23.01.09 Large Grant - Beyond metallicity: Exploiting the full POtential of CHemical elements (EPOCH) (ref. Laura Magrini).
 This work was supported by the Deutsche Forschungsgemeinschaft (DFG, German Research Foundation) – Project-ID 279384907 – SFB 1245, the State of Hessen within the Research Cluster ELEMENTS (Project ID 500/10.006). 
Finally, we thank an anonymous referee for very useful suggestions that improved the paper.

\bibliographystyle{aa} 
\bibliography{disk}

\end{document}